\documentclass[a4paper,twoside]{article}

\usepackage{epsfig}
\usepackage{subfigure}
\usepackage{calc}
\usepackage{amssymb}
\usepackage{amstext}
\usepackage{amsmath}
\usepackage{amsthm}
\usepackage{multicol}
\usepackage{pslatex}
\usepackage{tabularx}
\usepackage[capitalise]{cleveref}
\usepackage{dl4cv}

\begin{document}

\title{Detecting Bone Lesions in X-Ray Under Diverse Acquisition Conditions
}

\author{\authorname{Tal Zimbalist\sup{1} \qquad Ronnie Rosen \sup{1} \qquad Keren Peri-Hanania \sup{1} \qquad Yaron Caspi \sup{1} \qquad Bar Rinott \sup{2} \qquad Carmel Zeltser-Dekel \sup{2} \qquad Eyal Bercovich \sup{2} \qquad Yonina C. Eldar \sup{1} \qquad Shai Bagon \sup{1}}
\affiliation{\sup{1}Department of Computer Science \& Applied Mathematics, Weizmann Institute of Science, Rehovot, Israel.}
\affiliation{\sup{2}Medical Imaging Division, Rambam Health Care Campus, Haifa, Israel.}
}

\abstract{
The diagnosis of primary bone tumors is challenging, as the initial complaints are often non-specific. Early detection of bone cancer is crucial for a favorable prognosis. Incidentally, lesions may be found on radiographs obtained for other reasons. However, these early indications are often missed. In this work, we propose an automatic algorithm to detect bone lesions in conventional radiographs to facilitate early diagnosis. Detecting lesions in such radiographs is challenging: first, the prevalence of bone cancer is very low; any method must show high precision to avoid a prohibitive number of false alarms. Second, radiographs taken in health maintenance organizations (HMOs) or emergency departments (EDs) suffer from inherent diversity due to different X-ray machines, technicians and imaging protocols. This diversity poses a major challenge to any automatic analysis method.
We propose to train an off-the-shelf object detection algorithm to detect lesions in radiographs. The novelty of our approach stems from a dedicated preprocessing stage that directly addresses the diversity of the data. The preprocessing consists of self-supervised region-of-interest detection using vision transformer (ViT), and a foreground-based histogram equalization for contrast enhancement to relevant regions only. 
We evaluate our method via a retrospective study that analyzes bone tumors on radiographs acquired from January 2003 to December 2018 under diverse acquisition protocols. 
Our method obtains 82.43\% sensitivity at 1.5\% false-positive rate and surpasses existing preprocessing methods. 
For lesion detection, our method achieves 82.5\%  accuracy and an IoU of 0.69.
The proposed preprocessing method enables to effectively cope with the inherent diversity of radiographs acquired in HMOs and EDs.
}

\onecolumn 
\maketitle 
\normalsize 

\section{Introduction}
\noindent Malignant bone tumors are relatively rare and account for approximately 3\% of tumors in children and adolescents \cite{Jackson2016}. Osteosarcoma (OS) and the Ewing sarcoma family of tumors (ESFT) are the most common malignant bone tumors in children and adolescents~\cite{Jackson2016}. Bone sarcomas are ranked as the third leading cause of cancer death among humans up to 20 years old in the United States~\cite{Siegel2022}. Early diagnosis of a malignant bone tumor is of vital importance since it may not only increase the chance of survival, but also the possibility of performing a limb-sparing resection~\cite{Salom2021}. However, the nonspecific signs and symptoms in patients with primary bone tumors often lead to a delay in the diagnosis and initiation of treatment.

X-ray is the imaging modality frequently used for the initial evaluation of bone lesions~\cite{Mintz2014} since it is highly available and relatively inexpensive. However, many radiologists are not able to develop sufficient expertise to reliably identify and assess these lesions on radiographs due to their low prevalence in the population. Additionally, early indications of bone lesions may be incidentally missed, as radiographs taken in HMOs or EDs are usually obtained for other clinical reasons, and the radiologists examining the scan simply do not look for lesions in the scans.

In recent years, AI has gained popularity in diagnostic imaging to automatically recognizing complex patterns in imaging data and providing quantitative assessments of radiographic characteristics~\cite{Tang2020}, though its clinical applicability remained challenging.
There are two main challenges any such automatic method faces: The first is the low prevalence of lesions in scans, which dictates very high precision to avoid prohibitive false alarm rates. The second challenge is caused by the diverse data acquisition conditions. Different acquisition conditions, such as different hardware and imaging protocols results in images in different bit depth and variety of artifact~\cite{Bell2014}. These differences although insignificant for the human expert, pose a great challenge for any computerized analyzing methods which are in general very sensitive to these changes~\cite{Ng2021}.
In addition, each radiograph consists of a large background area which corrupts the image statistics. Therefore, classic contrast enhancement methods fail in producing high contrast detailed images.

In this work we overcome these challenges by directly addressing the diversity resulted by acquisition conditions; we explicitly enhance important anatomical features while eliminating background artifacts, thus extracting more knowledge from the limited dataset.
To summarize, in this work we noticed the diversity of clinical radiographs, which is exhibited in narrow histogram and poorly allocated dynamic range, thus low contrast. We introduce a simple foreground-aware histogram equalization (HE) to overcome this challenge and enhance the important signal in the scanned image. We further showcase the effectiveness of our method by developing a deep learning model for the classification and detection of bone lesions, as an example of such analyzing methods. We show how our preprocessing method significantly boosts the performance of our lesion detection model, given the same training and evaluation data. 

\begin{figure*}
    \centering
    \includegraphics[width=\linewidth]{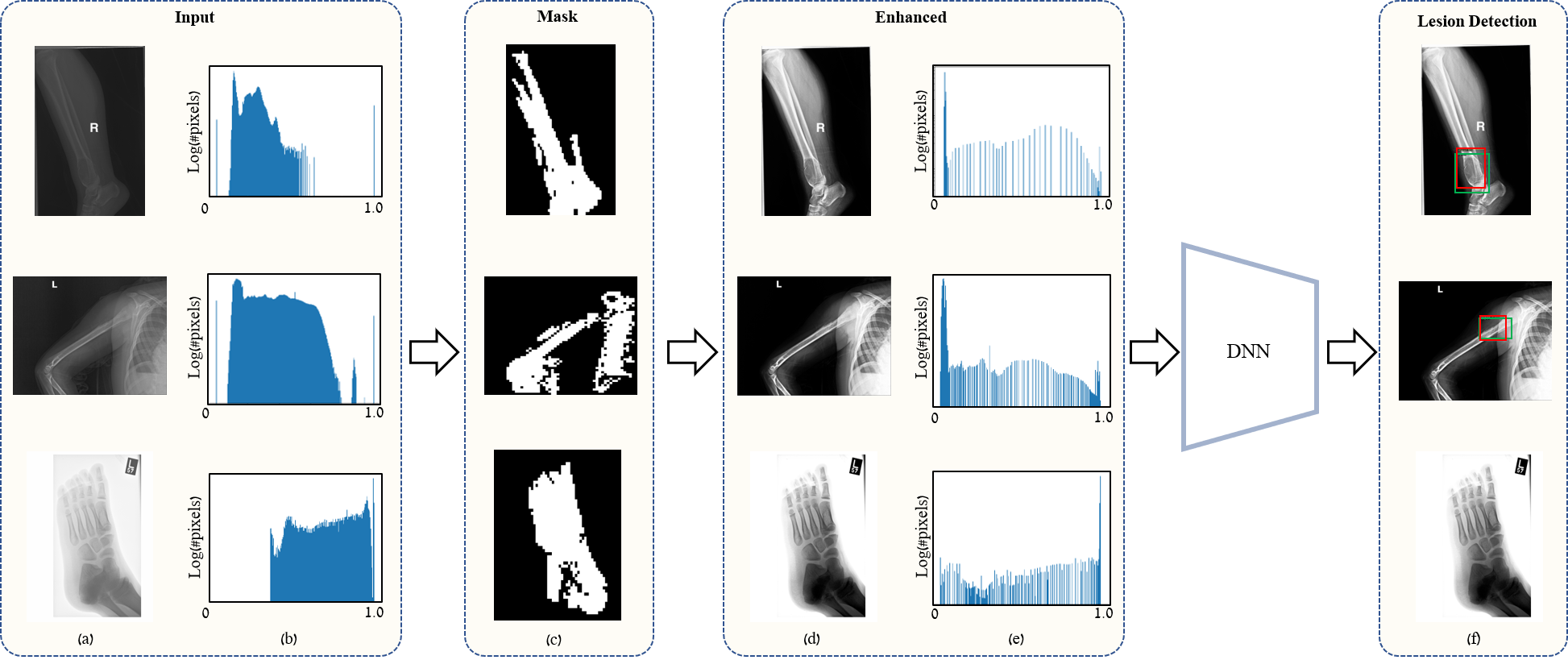}
    \caption{Overview: our method takes in low contrast images (a) with poorly allocated dynamic range (b) due to diverse acquisition conditions, and extracts ROI segmentation (c) using self-attention maps. We use the ROI masks to generate enhanced images (d),(e) that highlight the ROI and eliminate background artifacts. This process is applied to every image during training and in test time. The enhanced images are used as input for the deep neural network (DNN), that outputs bone lesions detection (f).}
    \label{fig:teaser}
\end{figure*}

\section{Related Work.}
\noindent
In this section, we discuss related work. First, we review previous studies of bone lesions analysis via deep learning algorithms. Next, we provide background on classic contrast enhancement methods, that are commonly used in medical imaging. Finally, we provide a brief overview of vision transformers related to the way we utilize them in this work.

\subsection{Bone Cancer Analysis on Radiographs.}
Several studies have used deep learning models to analyze bone tumors on radiographs. Yu He et al.~\cite{He2020} used a deep learning model to classify primary bone tumors in a multi-institutional dataset, however, the images were cropped by musculoskeletal radiologists to highlight the tumor before being inputted into the network, thus limiting the applicability of their method to be used for scanning arbitrary radiographs obtained for other clinical reasons. Von Schacky et al.~\cite{vonSchacky2021} used a multi-task deep learning model for simultaneous bounding box placement, segmentation, and classification of primary bone tumors on radiographs. However, they only considered radiographs of patients that had received a diagnosis of either benign tumors or malignant tumors, thus limiting clinical applicability in HMOs and EDs. Neither of the previous studies coped with the challenge of obtaining high precision while avoiding false alarms. In addition,~\cite{vonSchacky2021} ignores the diversity of the data while our method explicitly accounts for at least some aspects of the diversity, resulting in two benefits: reducing the diversity the model sees and better utilization and efficiency of the training data.

\subsection{Contrast Enhancement Methods.}
The information inherent in image histogram can be quite useful for different image processing applications, as it provides important image statistics. Existing contrast enhancement methods such as HE and contrast limited adaptive histogram equalization (CLAHE), use a transformation function based only on information available in the histogram of the input image, to achieve an effect of high contrast detailed image. HE modifies the pixels values in a way that the intensity level histogram of the equalized image spans a wider range of the intensity scale, thus resulting in enhanced contrast~\cite{Gonzalez2008}. HE is useful for increasing image’s global contrast, though it has limitations as it does not consider any high-level information in the image during the equalization process. Adaptive histogram equalization is a contrast enhancement method that demonstrated excellent results on both natural and medical images~\cite{Pizer1987}. It differs from HE in that it computes several histograms corresponding to different sections of the image, which makes it suitable for improving the local contrast. CLAHE~\cite{1571698601099987968} is a variant of adaptive histogram equalization in which the contrast amplification is limited to reduce the problem of noise over-enhancement which is typical to adaptive histogram equalization~\cite{Pizer1987}. Although, clipping level must vary with the imaging modality, body region imaged and imaging variables, thus limiting its robustness.
We use these methods as preprocessing baselines in our experiments, and we show that using our diversity-driven preprocessing method, the performance of an off-the-shelf lesion detector improve substantially.

\begin{figure*}[h]
    \centering
    \includegraphics[width=\textwidth,height=\textheight,keepaspectratio]{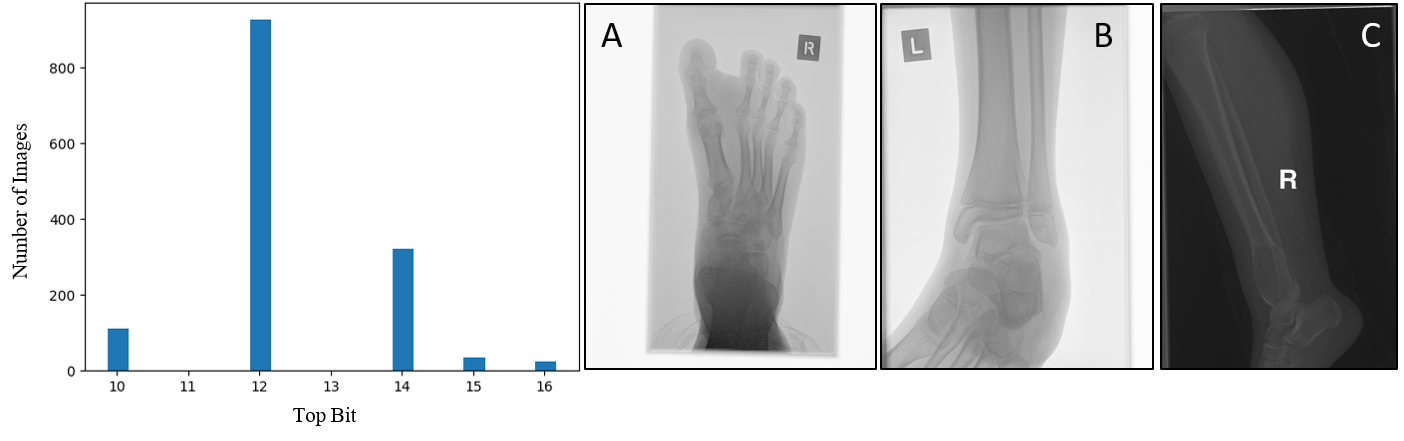}
    \caption{Dataset Characteristics. Left: data distribution. Right: low contrast images. A: 39 year-old male with a malignant tumor on the foot.  B: 12 year-old male with malignant tumor on the ankle. C: 13 year-old female with malignant tumor on the tibia.}
    \label{fig:Dataset Characteristics}
\end{figure*}

\subsection{Vision Transformers.}
Vision transformers (ViTs)~\cite{Dosovitskiy2020}, are recent popular deep learning architectures for image analysis, which often outperform state-of-the-art convolutional neural networks in terms of accuracy and computational efficiency. In ViTs, an image is processed as a sequence of non-overlapping patches and a class token that serves as the global representation of the image. The tokens are passed through $L$ Transformer layers where each layer is based on a self-attention mechanism~\cite{Vaswani2017} that computes the attention between patches. 
\par DINO-ViT~\cite{Caron2021} is a vision transformer that has been trained in a self-supervised manner using self-distillation approach with no lables. It learns powerful and semantically meaningful representations. The class token serves as the global representation of the image and the attention maps illustrates that the model automatically learns class-specific features that leads to self-supervised foreground segmentation. It has been shown in several applications that the class token attention map of trained DINO-ViT models provides good results in foreground detection. See e.g~\cite{amir2021deep}.
In this work we take advantage of this property and combine it with existing contrast enhancement method to obtain enhanced images while avoiding noise amplification and improve the performance of a simple object detector.

\begin{figure*}[h]
    \centering
    \includegraphics[width=\textwidth,height=\textheight,keepaspectratio]{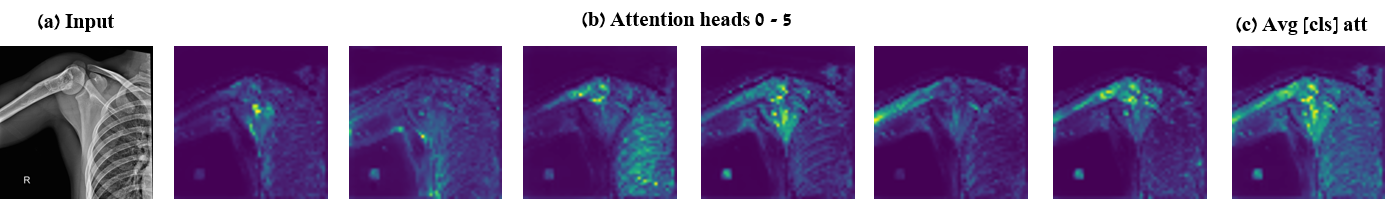}
    \caption{Attention heads and cls attention map extracted from the last layer of DINO-ViT/S-8. We show that even though DINO-ViT/S-8 was trained on natural images, it impressively detects the different organs in X-ray images. (radiograph of a 19-year-old male with a benign tumor at the shoulder).}
    \label{fig: Attention Heads}
\end{figure*}

\section{Method}
This retrospective study was approved by the institutional review boards (IRB) of both Rambam Health Care Campus (HCC study approval number RMB-0663-18) and by the Weizmann Institute of Science IRB committee.



\par The inherent diversity of radiographs degrades the performance of lesion detection algorithms. Our method directly addresses the inherent diversity of radiographs caused by acquisition conditions as exhibited by poorly allocated dynamic range. We apply foreground-based histogram equalization (HE) in a way that only relevant pixels are considered in the enhancement process. In this way, we emphasize the important signal only, which is important when dealing with limited training data. Foreground masks are generated in an automatic manner that demonstrates robustness to different image intensity acquisition characteristics and imaged organs. Figure~\ref{fig:teaser} shows an overview of our method. We propose a method that takes diverse dataset as an input (a),(b), and extracts foreground masks in a self-supervised manner via self-attention (c). We use those mask to perform histogram equalization (HE) based on the distribution of relevant pixels only, resulting in enhanced images (d),(e). The enhanced images are used as the input of an off-the-shelf object detection algorithm that outputs lesion detection as seen in Figure ~\ref{fig:teaser} (f). We perform the enhancing process to both training and test datasets. Using this diversity-driven method we benefit twice: we explicitly address the diversity resulted by acquisition conditions, thus extracting more knowledge from the limited training data to achieve high recall at low false-positive rate.
\par In this section we first provide background on self-supervised foreground detection using ViT's. Next, we describe our proposed enhancement method applied to both train and test sets. Finally, we detail the automatic lesion detection algorithm and evaluation metrics.

\subsection{Foreground Detection Using DINO-ViT}
\label{sec: forground detection using DINO-ViT}

As can be seen in Figure~\ref{fig:Dataset Characteristics}, each radiograph consists of a large background area which corrupts the image statistics. Thus, we use the meaningful representations DINO-ViT learns to separate the object from the background.
We show that, surprisingly, simply using ImageNet pretrained DINO-ViT to extract the attention maps leads to successful foreground detection on a completely different domain such as medical imaging. (See Figure~\ref{fig: ROI Segmentation} and Figure~\ref{fig: Enhancing Process}). Figure~\ref{fig: Attention Heads} illustrates the attention heads and the cls attention of a 19 year-old male with benign tumor at the shoulder. The cls attention detects the foreground using an off-the-shelf model and pretrained weights, even though no training or finetuning was involved in the process.
\par Figure~\ref{fig: Attention Heads} also demonstrates that it is possible to detect specific organs by applying mathematical manipulations to different heads instead of the cls attention, as each head corresponds to different body part of the patient.

\subsection{Enhancing Process}
\begin{figure*}
    \centering
    \includegraphics[width=\textwidth,height=\textheight,keepaspectratio]{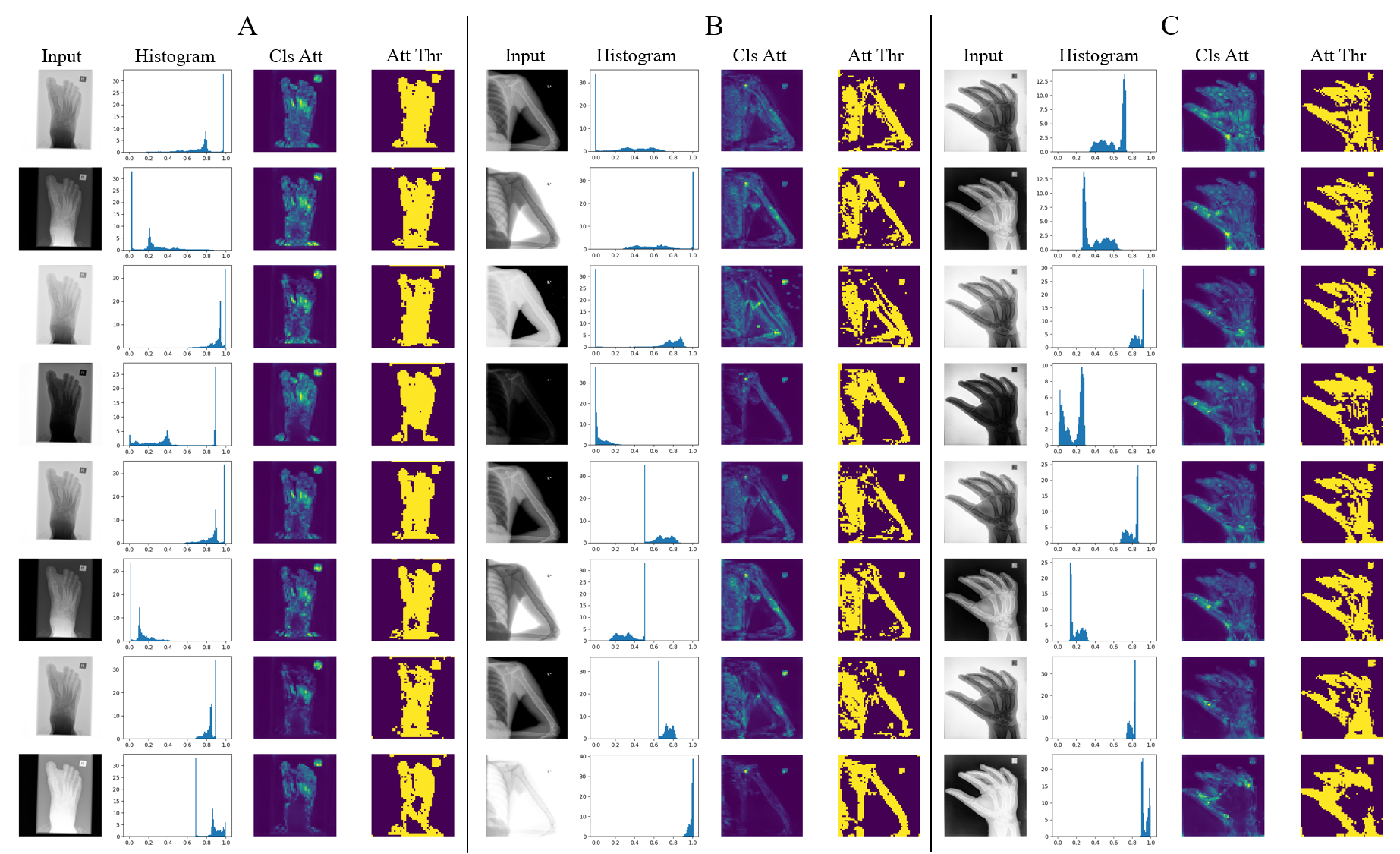}
    \caption{ROI segmentation. From left to right: input image, histogram, class token attention map, thresholded attention map (70th percentile). We artificially demonstrate the robustness of DINO-ViT to diverse input images (first row) by using different augmentations-contrast inversion, scaling, shifting, etc. (rows 2-8). Images A-C illustrates DINO-ViT robustness to different scanned organs: A: 39 year-old male with a malignant tumor in his foot.  B: 26 year-old male with benign lesion in his shoulder. C: 23 year-old male with benign lesion in his hand.}
    \label{fig: ROI Segmentation}
\end{figure*}

\begin{figure*}
    \centering
    \includegraphics[width=\textwidth,height=\textheight,keepaspectratio]{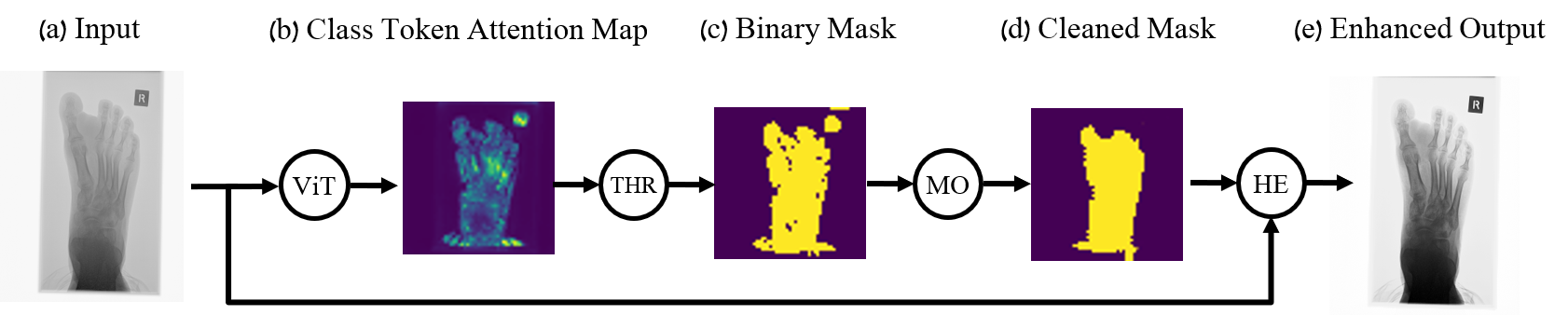}
    \caption{Full contrast enhancement process: Input images (a) are fed to DINO ViT-S/8 to obtain attention maps (b). Foreground binary masks (c) are generated by employing 70th percentile thresholding on the attention maps. Blobs are removed (d) by applying morphological operations (MO). Finally, HE is applied on the input images based on the distribution of the foreground intensity levels only to yield enhanced images (e). The process is applied to each input X-ray scan.}
    \label{fig: Enhancing Process}
\end{figure*}

Contrast enhancement methods adjust the image contrast while considering the intensity level distribution of the entire image. Contrast enhancement based on the region of interest (ROI) intensity levels only, especially in images where the ROI is relatively small, can improve the enhancement process while avoiding noise amplification. 
Figure~\ref{fig: ROI Segmentation} illustrates three input images and their corresponding poorly allocated dynamic range resulted by acquisition conditions. For each image (A-C) in Figure~\ref{fig: ROI Segmentation}, the image in the top row is the original radiograph given by the clinical collaborator, while the other rows artificially demonstrates non-ideal dynamic range conditions. We show that the class token attention map not only detects the object successfully in standard radiographs (as discussed in section~\ref{sec: forground detection using DINO-ViT}), but also demonstrates robustness to diverse imaging conditions including contrast inversion, shifting, and scaling (Figure~\ref{fig: ROI Segmentation} rows 2-8). It is also robust to the scanned organs as each input image (Figure~\ref{fig: ROI Segmentation} A-C) represents different organ. 
Finally, we take the foreground mask and apply histogram equalization based on the distribution of the ROI intensity levels. The full enhancing process is illustrated in Figure~\ref{fig: Enhancing Process}. This process is applied to each input X-ray scan (namely, both train and test sets), resulting in 16-bit enhanced images.

The proposed enhancing method is quite general and can be applied for different medical imaging applications and tasks. We demonstrate our method on two tasks – abnormality classification and detection of bone tumors on radiographs.

\section{Implementation Details}
\label{sec:implementation_details}

The following sections describe the general characteristics of our clinical dataset and the implementation details of an off-the-shelf algorithm used for both bone tumor classification and detection tasks.

\subsection{Data Characteristics}
The clinical dataset was acquired at the Medical Imaging Division, Rambam Health Care Campus, Haifa, Israel. 
As shown in Table~\ref{tab:patient-characteristics}, this retrospective study analyzed bone tumors on 1421 radiographs (mean age $23$ years $\pm 8$ [standard deviation]; $1001$ men). This retrospective study analyzed bone tumors on 1421 radiographs obtained between January 2003 and December 2018, 973 of which are normal images, with no pathology or any clinical finding, and 448 abnormal images that include benign or malignant lesions. Radiographs included in this study were acquired from patients aged 5-40 years. Radiographs with poor image quality were excluded. In addition, postoperative scans, chest and abdominal scans, and scans that were non-relevant for diagnosis were also excluded. Participants were identified based on ICD code in the electronic medical record. Imaging data was reviewed by two senior radiologists and one radiology resident to establish ground truth.

\begin{table*}

    \caption{\textbf{Radiographs Characteristics:} Age expressed as mean $\pm$ standard deviation.}
    \centering
    \begin{tabular}{|c|c|c|c|}
        \hline
        \textbf{Characteristic} & \textbf{Overall} (n=1421) & \textbf{Training Set} (n=1148) & \textbf{Test Set} (n=273) \\
        \hline
        Age [y] & $23\pm8$ & $23\pm9$ & $24\pm8$ \\
        \hline
        Male & 1001 & 812 & 189 \\
        \hline
        Normal & 973 & 774 & 199 \\
        \hline
        Abnormal & 448 & 374 & 74 \\
        \hline
        Benign & 137 & 122 & 15 \\
        \hline
        Malignant & 311 & 252 & 59 \\
        \hline
    \end{tabular}
    \label{tab:patient-characteristics}
    
\end{table*}

The data was acquired over a decade with a variety of X-ray machines and imaging protocols, resulting in a diverse dataset in terms of image characteristics. Image quality is determined by several factors including spatial resolution, patient movement and the technician which has an important role in maintaining good image quality~\cite{Byerly2022}. Figure~\ref{fig:Dataset Characteristics} (left) shows that the dataset consists of images with different bit depths – representing significant variations in the captured dynamic range. Moreover, in some cases, dynamic range is not fully utilized which means that the dataset consists of low contrast images as can be seen in Figure~\ref{fig:Dataset Characteristics} (right). In addition, the dataset not only varies in terms of intensity characteristics but also heterogeneously in terms of the scanned organs.

\subsection{Data Preparation}
The radiology team classified each tumor-containing radiograph as depicting benign or malignant tumors and additionally performed bounding box placement of the tumors. Images with no pathology were classified as normal images. The dataset was split into training and testing sets, composed of 80\% (1148/1421) and 20\% (273/1421) respectively. The division was carried out to avoid any overlap of patients between the training and testing datasets, a crucial step in maintaining the validity of our study's findings.

\subsection{DINO-ViT for ROI Segmentation}
We use ImageNet pre-trained DINO ViT-S/8 with overlapping patches (stride=4)~\cite{amir2021deep} to extract the self-attention maps from the last transformer layer. We employ 70th percentile thresholding to the average attention map to generate ROI binary mask. Morphological operations were used to clean up blobs in the binary mask and the largest object, which represents the scanned organ, was extracted using connected components.

\subsection{Lesion Detection}
We train an off-the-shelf object detection algorithm; we use Detectron2’s~\cite{wu2019detectron2} implementation of Faster R-CNN~\cite{Ren2015} model using R101-FPN as a base model. Both pixel-level and spatial-level augmentations were applied using Albumentations library~\cite{Buslaev2020}. The loss functions used for bounding box placement and classification are regression loss and cross-entropy loss respectively. The model outputs one or more bounding boxes, where each bounding box has a class label and a confidence score. Output with no bounding boxes is considered as normal image with no bone lesions.

\section{Results}
\noindent
\subsection{Statistical Analysis}
Since some benign bone lesions may require treatment (complete local excision/curettage) and should be observed by a specialist, we evaluate the model by binary classification into abnormal (malignant or benign) and normal radiographs. For model evaluation, we use the receiver operating characteristic (ROC) curve and the area under the curve (AUC). For bounding box placement, we used accuracy (detected instances out of total instances) and intersection over union (IoU) similar to~\cite{vonSchacky2021}. Correct placement of bounding box assumed for IoU$>$0.5. We evaluate on a set of 273 radiographs including benign and malignant tumors and normal images with no lesions or any other abnormality. All statistical analyses were performed by using scikit-learn, version 0.22.1~\cite{Buitinck2013}. 

\subsection{Baselines}
We compared the performance of an off-the-shelf object detection algorithm using different preprocessing methods. The first one is a naïve conversion of our diverse dataset to 8-bit images by simply dividing each image by its bit depth and multiplying by 255, similar to~\cite{vonSchacky2021}. Other preprocessing methods that we examined included applying 16-bit histogram equalization (HE) and contrast limited adaptive histogram equalization (CLAHE) on the 8-bit images. For each of these baseline preprocessing methods, we trained a lesion detection network using the configuration described in Sec.~\ref{sec:implementation_details}, and compared the performance of the resulting detectors.

\subsection{Abnormality Classification Results}
\begin{figure}[h]
    \centering
    \includegraphics[width=8.5cm,height=\textheight,keepaspectratio]{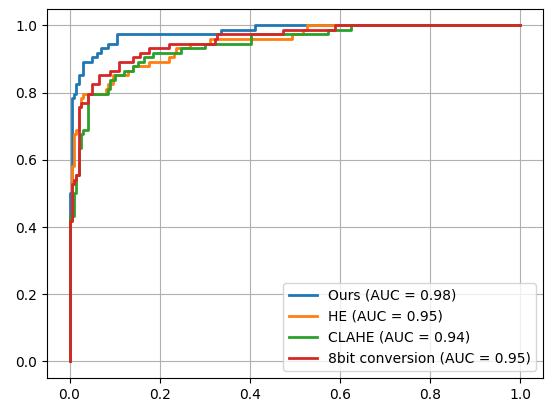}
    \caption{ROC curve for normal / abnormal classification}
    \label{fig: ROC Curve}
\end{figure}

The ROC curve is provided in Figure~\ref{fig: ROC Curve}, showing the trade-off between the false positive rate (FPR) and the true positive rate (TPR) for different classification threshold values. Since bone lesions incidence among the population is low~\cite{Siegel2022}, we focus on the low FPR area of the ROC curve (in this case 1.5\%) to avoid prohibitive false alarm rates. Table~\ref{tab:classification-results} gives an overview of the performance of the off-the-shelf object detector on the test set at 1.5\% FPR. Data in brackets represents 95\% confidence interval (CI). We allow only 4 false alarms across the entire test set; under this constraint, the deep learning model obtains 82.43\% sensitivity with our diversity-aware preprocessing method, compared to 68.91\% and 55.4\% sensitivity with HE and CLAHE respectively. The deep learning model also obtains 55.4\% sensitivity at zero FPR with naïve conversion to 8-bit images. Both na\"ive conversion and CLAHE include int8 quantization, meaning that information loss due to data quantization degrades the model performance. The AUC for our method is 0.98, which is the highest AUC out of all preprocessing baselines that were considered in this study. Our method surpasses all others across the board as our method efficiently overcomes the diversity in the data using the proposed preprocessing phase.

\begin{table}
    \caption{\textbf{Classification results:} Metrics comparison of Faster R-CNN trained with various preprocessing methods. Data in brackets are 95\% CIs}
    
    \centering
        \begin{tabularx}{\columnwidth}{c|c|c}
            \hline
            Preprocessing Method & AUC & Sensitivity @1.5\% \\ 
            & & FPR \\
            \hline
            \textbf{Ours} &0.98 & 82.43\% [77.9, 86.9]  \\
            HE & 0.95 & 68.91\% [63.4, 74.4]  \\
            CLAHE & 0.94 & 55.40\% [49.5, 61.3]  \\
            8bit Conversion & 0.95 & 55.40\% [49.5, 61.3] \\
            \hline
        \end{tabularx}
    \label{tab:classification-results}
\end{table}

\subsection{Tumor Detection Results}
Table~\ref{tab:detection-results} gives an overview of the detection performance comparison. Our model placed 82.5\% (66 out of 80 instances; 95\% CI: 74.17, 90.83) of the bounding boxes correctly (IoU$>$0.5) and demonstrates an IoU of 0.69±0.11. Our method surpassed the baselines in detection accuracy as both HE and CLAHE obtained 81.25\% while naive 8-bit conversion obtained only 77.5\% accuracy. In terms of IoU, naive 8-bit conversion obtains the highest result (0.7±0.10) out of all the preprocessing methods.

\begin{figure*}
    \centering
    \includegraphics[width=\textwidth,height=\textheight,keepaspectratio]{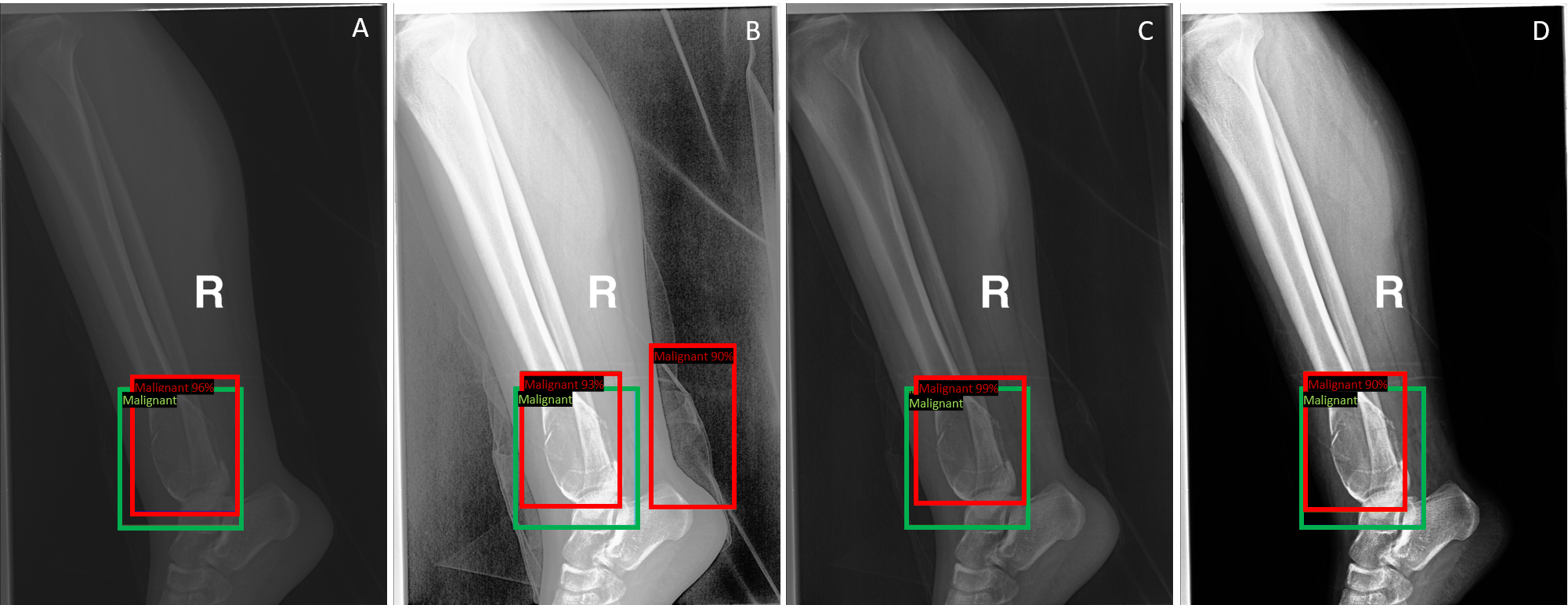}
    \caption{Detection examples of a 13-year-old girl  with a malignant tumor with different preprocessing methods. A-D: 8 bit naïve conversion, HE, CLAHE, our diversity aware method. Using HE, the model detects background artifact as a malignant tumor with high confidence. Our method (D) highlights the ROI while eliminating those background artifacts. Note: ground truth (GT) presented as a green bounding box with class name at its top left corner, prediction is presented as a red bounding box with class name and confidence score at the top left corner.}
    \label{fig: Detection Example 13y/o}
\end{figure*}

\begin{figure*}
    \centering
    \includegraphics[width=\textwidth,height=\textheight,keepaspectratio]{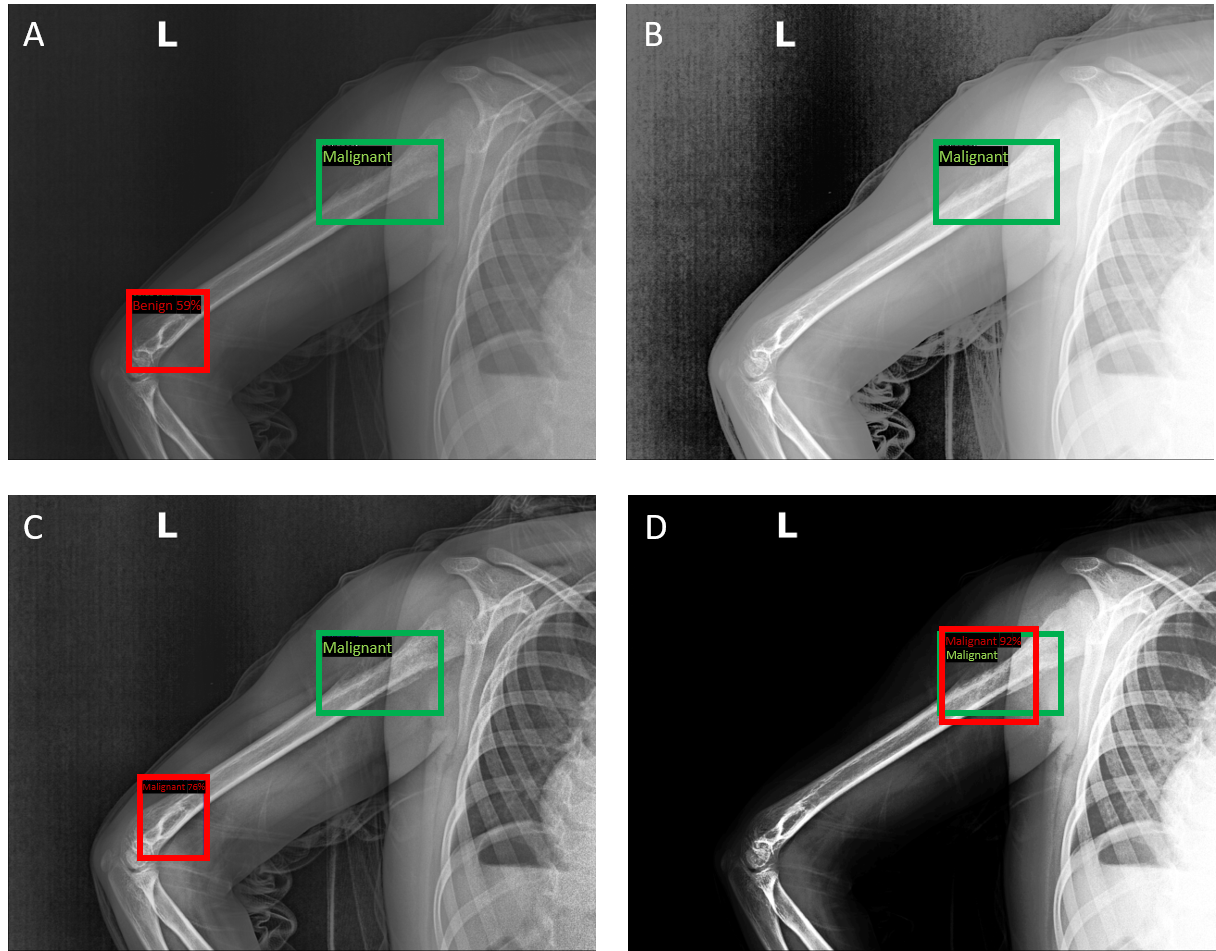}
    \caption{Detection examples of an 8-year-old girl with a malignant tumor with different preprocessing methods. A-D: 8 bit naïve conversion, HE, CLAHE, our diversity aware method. Using 8bit naïve conversion and CLAHE, the model incorrectly detects benign and malignant tumors respectively with high confidence, while using our method the model managed to successfully detect (IoU$>$0.5) the tumor. Note: ground truth (GT) presented as a green bounding box with class name at its top left corner, prediction is presented as a red bounding box with class name and confidence score at the top left corner.}
    \label{fig: Detection Example 8y/o}
\end{figure*}

Figure~\ref{fig: Detection Example 13y/o} demonstrates detection results of a 13-year-old female with a malignant tumor of the tibia. The model placed the bounding box correctly (IoU$>$0.5) and classified the tumor correctly as a malignant tumor in all preprocessing methods. However, using HE the model detects background artifacts as a malignant tumor with confidence 90\%, meaning that in some cases histogram equalization can overamplify noise by enhancing unwanted artifacts instead of enhancing the region of interest. Our method eliminates background artifacts caused by objects external to the patient (e.g clothing) in addition to region of interest contrast enhancement. Using our method, we lose some of the patient’s soft tissue, but it does not affect bone tumor detection. 
Figure~\ref{fig: Detection Example 8y/o} demonstrates detection results of an 8-year-old girl with a malignant tumor of the humerus. As can be seen in Figure~\ref{fig: Detection Example 8y/o}, the model successfully detects the tumor only using our diversity-aware preprocessing method. Using naïve 8bit conversion and CLAHE, the model detects benign and malignant tumors respectively with confidence greater than 50\% incorrectly, as the bounding box placement is completely off. Our method can be fitted to different clinical applications by setting different kernels for the morphological operations or using other attention map to extract the ROI segmentation; averaging specific attention heads, exploring the attention heads in the different transformer layers.

\begin{table*}
    \caption{\textbf{Lesion detection results:} Note: correct placement of bounding box assumed for IoU$>$0.5. Data for accuracy is expressed as detected instances out of total instances. Data listed in brackets are 95\% confidence intervals (CIs). IoU expressed as mean±standard deviation.}
    \centering
        \begin{tabular}{|c|c|c|c|c||c|c|}
            \hline
            Method & Ours &HE &CLAHE & 8bit& v. Schacky et al.\cite{vonSchacky2021} & v. Schacky et al.\cite{vonSchacky2021}\\ 
             & & & & & Internal test set& External test set\\ 
            \hline
            Accuracy  &82.50\% &81.25\% &81.25\% &77.50\% &65.0\% &59.50\% \\
            & (66/80) & (65/80) & (65/80) & (62/80) & (91/140) & (66/111) \\
            & [74.1, 90.8]&[72.7, 89.8]&[72.7, 89.8]&[68.3, 86.6]&[57.1, 72.9]&[50.3, 68.6]\\
            \hline
            IoU& 0.69±0.11& 0.69±0.09& 0.68±0.10& 0.70±0.10& 0.54±0.32& 0.52±0.34\\
            \hline
        \end{tabular}
    \label{tab:detection-results}
\end{table*}

A previous study by Von Schacky et al.~\cite{vonSchacky2021} ignores the diversity of the data and trained a deep learning model using the same used backbone (ResNet101) only on 8-bit images. They used a different dataset which included both internal and external datasets (data from a different university hospital in the same country) and obtained 65\% and 59.50\% detection accuracy and IoU of 0.54±0.32 and 0.52±0.34 respectively. In contrast to~\cite{vonSchacky2021}, our method enables training on 16-bit images and therefore avoids quantization error resulting from conversion to 8-bit.

\subsection{Runtime Analysis}
\label{sec:runtime}
We have conducted a comprehensive assessment of the computational demands of each stage in our enhancing process. The analysis results in Table~\ref{table:gpu_performance} reveals the model's runtime performance for key preprocessing steps— cls token attention map extraction,  cleaned mask extraction, and contrast enhancement—across different GPUs. These metrics are critical, considering the necessity for rapid processing in a clinical context.

The Nvidia A100 GPU, representing the pinnacle of current technology, processes feature extraction within 1.8667 seconds for the 'Small' configuration of the Vision Transformer (ViT), and 2.1323 seconds for the 'Base' configuration. This demonstrates the model's capacity to integrate seamlessly into clinical workflows without substantial delays. The Tesla V100 and RTX2080 GPUs also deliver competitive runtimes, ensuring that the model's deployment is feasible even in clinical settings with varying levels of computational resources.

\begin{table*}[htbp]
\caption{Runtime performance comparison for various enhancement process stages—cls token attention extraction, mask extraction, and contrast enhancement—across different GPU models and configurations. The reported runtimes are in seconds for an input image size of 2000x1000 pixels.}
\begin{center}
\begin{tabular}{|c|c|c|c|c|}
\hline
\textbf{GPU} & \textbf{ViT Type} & \textbf{Attn map extraction} & \textbf{Mask extraction} & \textbf{Contrast enhancement} \\
 & & \textbf{Runtime [sec]} & \textbf{Runtime [sec]} & \textbf{Runtime [sec]} \\
\hline
Nvidia A100 & Small & 1.8667 & 0.0375 & 0.0284 \\
 & Base & 2.1323 & 0.0918 & 0.0284 \\
Tesla V100 & Small & 2.1397 & 0.0382 & 0.0364 \\
 & Base & 2.1745 & 0.1142 & 0.0336 \\
RTX2080 & Small & 1.7547 & 0.0597 & 0.0285 \\
 & base & 2.8636 & 0.1548 & 0.0279 \\
\hline
\end{tabular}
\label{table:gpu_performance}
\end{center}

\end{table*}

\begin{table}[htbp]

\caption{Detectron2 inference runtime performance across different GPU nodes. Detectron2 configuration is described in Section~\ref{subsec:configuration}. The reported runtimes are in seconds for an input image size of 2000x1000 pixels.}
\begin{center}
\begin{tabular}{|c|c|}
\hline
\textbf{GPU} & \textbf{Detectron2 Inference Time} \\
\textbf{} & \textbf{Runtime [sec]} \\
\hline
Nvidia A100 & 2.9473 \\
Tesla V100 & 3.9367 \\
RTX2080 & 4.2338 \\
\hline
\end{tabular}
\end{center}
\label{table:detectron2_gpu_performance}

\end{table}

Detectron2 inference times range from 2.9473 to 4.2338 seconds for an input image size of 2000x1000 pixels as shown in Table~\ref{table:detectron2_gpu_performance}. The overall runtime of our method showcases its adaptability to varying hardware configurations while maintaining rapid processing speeds. This versatility ensures that our approach remains applicable across a spectrum of clinical settings, where computational resources may vary, without necessitating extensive infrastructure upgrades. Together, these results affirm the feasibility of integrating our methodology into real-world clinical workflows.

\section{Discussion}
\label{sec:discussion}

\begin{quotation}
\noindent \textit{``It turns out [that when] you take […] that same AI system, […] and the technician uses a slightly different imaging protocol, that data drifts to cause the performance of AI system to degrade significantly"}\par
\begin{flushright}
Andrew~Ng~\cite{Ng2021}
\end{flushright}
\end{quotation}

In this work, we directly addressed the issue of data drift caused by different acquisition conditions (e.g imaging protocols and scanning machines). We introduced a method that explicitly address this issue and gain a robust framework capable of processing data from different clinical sites and imaging equipment to achieve high sensitivity while maintaining low false-positive rate, thus making it suitable for clinical setups. Our method allows treating images with different bit depth in a unified manner while utilizing the full dynamic range. Explicitly enhancing the ROI makes the finetuning process of the deep learning model efficient and robust to acquisition parameters. We showcase the effectiveness of our method by developing a deep learning model for the classification and detection of primary bone tumors. Using the same analysis algorithm we show how our preprocessing method significantly boost performance compared to existing enhancement methods.

Our study has several limitations. First, the model was not trained to detect other diseases such as fractures or osteoporosis. To make it clinically applicable it should be trained on a dataset that consists of radiographs with other abnormalities in addition to bone lesions. Second, in the general population, benign tumors are more common than malignant ones, yet our dataset contained a smaller number of benign bone tumors compared to malignant bone tumors, indicating potential bias. Third, our study did not consider patient age. Considering age may provide additional information to the model and improve its performance.

In conclusion, a deep learning model for detecting bone lesions can be beneficial in HMOs and EDs for analyzing radiographs that are not usually assessed by radiologists. The detection performance of the model surpassed previous study by Von Schacky et al.~\cite{vonSchacky2021}. The proposed method improves the robustness of the deep learning model and its ability to generalize across medical images acquired under diverse conditions and protocols, as the classification performance using this method surpassed that of other preprocessing methods. A framework that extracts more knowledge from the data is extremely important, especially in domains with lack of training data such as medical imaging. In a future study the proposed method should be demonstrated on different tasks to inspect its impact on deep learning algorithms performance.

\paragraph{Acknowledgement.}
Funding: This research was supported by the Israel Ministry of Health (grant 3-15891), the Carolito Stiftung, and NVIDIA's Applied Research Accelerator Program.
Dr. Bagon is a Robin Chemers Neustein AI Fellow.

\bibliographystyle{IEEEtran}
{\footnotesize
\bibliography{bone_lesions}}

\end{document}